# Spatial Differences Analysis of Nodal Frequency Considering Frequency-Voltage Coupling

Xiaodan Li, Donghai Zhu, *Senior Member, IEEE,* Zining Liu, Zhu Guo,
Jiabing Hu, *Senior Member, IEEE,* and Hong Rao, *Fellow, IEEE*

*Abstract*—**The rising penetration of inverter-based resources (IBRs) brings rich control dynamics that can amplify spatial differences in nodal frequency through frequency-voltage (F-V) coupling. Existing analyses often neglect voltage dynamics or only examine the aggregate frequency, leading to overoptimistic evaluation of frequency security. To address these drawbacks, this paper develops a unified model of synchronous generators (SGs), grid-following (GFL)-IBRs, and grid-forming (GFM)-IBRs based on the magnitude and phase dynamics of equivalent internal electromotive force (EMF). Integrating these source models with network equations yields a sparse implicit closed-loop model, which fully captures the device-network F-V coupling without transfer-matrix inversion. Then, the obtained nodal responses are decomposed into common and differential modes. Using this decomposition, coupling indices and pole-residue analysis reveal how F-V coupling can amplify spatial differences in nodal frequency and how source configuration and parameters modify this effect. Finally, case studies on the modified IEEE 39-bus and 118-bus test systems validate the proposed model and demonstrate its computational efficiency.**

## I. INTRODUCTION

THE increasing integration of inverter-based resources (IBRs) introduces diverse synchronization and control dynamics, posing new challenges to power-system frequency security [1]. In particular, nodal frequency responses may exhibit pronounced spatial differences [2]-[4]. However, the widely used center-of-inertia (COI) frequency and average system frequency (ASF) primarily characterize aggregate frequency motion and cannot adequately capture these local variations [5], [6]. Such variations may affect underfrequency load shedding [7] and frequency-constrained unit commitment [8]. Nodal frequency analysis is therefore essential for assessing local frequency risks and guiding frequency support allocation.

To address the limitations of aggregate frequency measures, existing studies have developed methods for analyzing spatial variations in nodal frequency through disturbance propagation, frequency mapping, and response decomposition. Electromechanical-wave models describe how disturbances propagate across the network using a continuum approximation [9]. However, this approximation offers limited representation of the distinct control dynamics of IBRs. The frequency-divider method relates nodal frequencies to synchronous-machine rotor speeds [10], but its original formulation does not explicitly capture voltage regulation and converter synchronization dynamics. From a modal perspective, analytical decomposition separates global and local frequency components [5], while characteristic-subsystem analysis identifies common and differential modes [11]. Both approaches provide insights into aggregate frequency behavior and spatial differences in nodal responses. Nevertheless, these decompositions generally employ simplified device models and neglect frequency-voltage coupling.

Such modeling simplifications become increasingly restrictive under high renewable penetration and weak-grid conditions [12], [13]. In such systems, source synchronization, power control, and voltage regulation interact through the network, forming a closed loop that couples frequency and voltage dynamics. Neglecting these interactions may lead to inaccurate assessments of local frequency excursions and spatial frequency differences. Previous studies have investigated frequency-voltage and active-reactive power coupling in grid-forming (GFM) converters [14], [15], as well as voltage-dependent frequency responses of grid-following (GFL) and GFM devices [16]. Studies based on internal electromotive force (EMF) models have further examined how control dynamics affect the frequency responses of doubly fed induction generators [17]-[19]. Interactions between different converter types have also been investigated [20]. However, these studies mainly focus on device-level dynamics and converter interactions, leaving their effects on spatial variations in nodal frequency insufficiently quantified.

At the system level, several studies have incorporated voltage dynamics into frequency analysis. Complex frequency relates the dynamics of bus-voltage magnitude and phase angle to complex-power variations [21], primarily characterizing local bus behavior. The complex frequency divider further establishes the relationship between nodal voltage dynamics and device current injections [22]. However, these formulations do not directly quantify how F–V coupling affects spatial differences in nodal frequency responses. Voltage-influenced common-mode frequency explicitly captures the effects of voltage dynamics on aggregate frequency through the F-Q-V-P coupling pathway [23], extending conventional frequency analysis. However, its globally coupled voltage term increases computational complexity, requiring network partitioning that introduces approximation errors. These limitations motivate a unified framework for quantifying the effects of device–network F–V coupling on spatial differences in nodal frequency, with computational efficiency.

To address these challenges, this paper develops a unified sparse implicit device-network closed-loop model accounting

This work was supported by the National Natural Science Foundation of China under Grant U25B6019. *(Corresponding author: Donghai Zhu).*

for source synchronization, voltage regulation, and network power coupling. The magnitude and phase dynamics of equivalent internal EMFs provide a common representation of different sources. Their interconnection with network equations preserves the complete F-V coupling without transfer-matrix inversion and maintains matrix sparsity for efficient computation. The main contributions are as follows:

1) A unified model incorporating the magnitude and phase dynamics of internal EMF preserves device-network F-V coupling, improving frequency and voltage prediction accuracy for SG, GFL, and GFM resources.

2) Frequency decomposition, combined with pole–residue analysis and coupling indices, provides a framework for assessing how F-V coupling, source configuration, and system parameters influence spatial frequency differences.

3) The sparse implicit formulation preserves network sparsity without transfer-matrix inversion or passive-bus elimination, enabling efficient computation and model reuse for repeated disturbance analyses.

The remainder of this paper is organized as follows. Section II develops the unified source models and sparse implicit device-network closed loop. Section III presents frequency decomposition, pole-residue characterization, and coupling-effect indices. Section IV examines coupling mechanisms and the effects of source configuration and control and network parameters. Section V validates the model, analytical findings, and computational performance through case studies. Section VI concludes the paper.

## II. Unified Sparse Implicit Device-Network Closed-Loop Model

This section considers a power system comprising SGs, GFL resources, and GFM resources to analyze nodal frequency responses under F-V coupling. Source-port models and network equations are first established and then combined into a sparse implicit closed-loop model. This model provides the basis for frequency decomposition and coupling analysis in Section III.

### A. Unified Implicit Port Model for Generation Devices

In a common synchronous reference frame, the equivalent internal EMF and terminal voltage of source $i$ are

$$\mathcal{E}_i = E_i e^{j\theta_{E,i}}, \mathcal{V}_i = V_i e^{j\theta_i}, \delta_i = \theta_{E,i} - \theta_i \tag{1}$$

where $E_i$ and $\theta_{E,i}$ are the internal-EMF magnitude and angle, $V_i$ and $\theta_i$ are the terminal-voltage magnitude and angle.

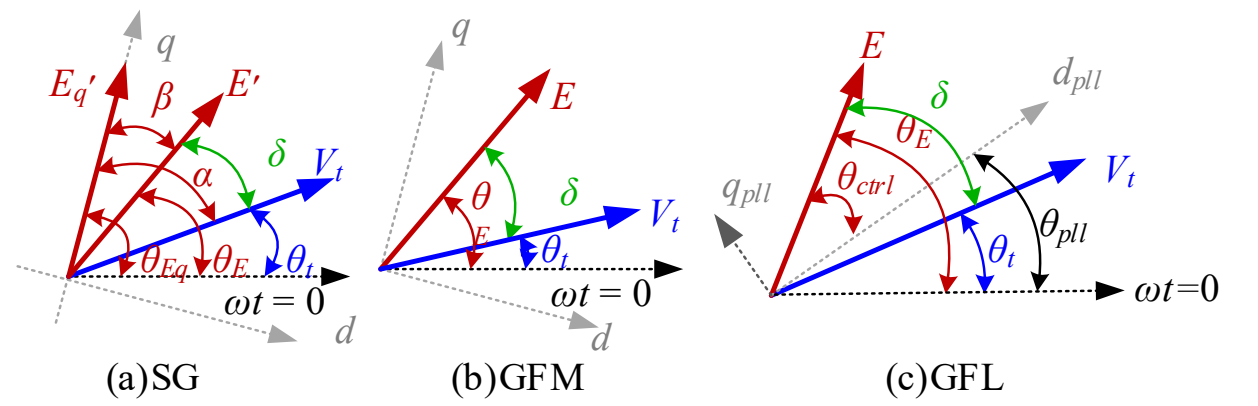

**Fig. 1.** Internal-EMF and terminal-voltage phasor relations. (a) SG. (b) GFM. (c) GFL.

Fig. 1 illustrates the internal-EMF and terminal-voltage phasor relationships for SG, GFM, and GFL sources. Define the port perturbations as $\Delta \boldsymbol{p}_i = [\Delta P_i, \Delta Q_i]^T$ and $\Delta \boldsymbol{x}_i = [\Delta\theta_i, \Delta v_i]^T$, where $\Delta P_i$ and $\Delta Q_i$ are positive for injection into the network, $\Delta v_i = \Delta V_i / V_{i0}$, and $\Delta\varepsilon_i = \Delta E_i / E_{i0}$. The subscript 0 denotes the operating point. Linearizing the terminal power relations gives

$$\begin{bmatrix} \Delta P_i \\ \Delta Q_i \end{bmatrix} = \begin{bmatrix} K_{P\delta,i} & K_{P\varepsilon,i} & K_{Pv,i} \\ K_{Q\delta,i} & K_{Q\varepsilon,i} & K_{Qv,i} \end{bmatrix} \begin{bmatrix} \Delta\delta_i \\ \Delta\varepsilon_i \\ \Delta v_i \end{bmatrix} \tag{2}$$

where $K_{ab,i} = \partial a_i / \partial b_i |_0$, $a \in \{P, Q\}$, and $b \in \{\delta, \varepsilon, v\}$.

Equation (2) relates terminal-power perturbations to changes in the internal EMF magnitude, its angle relative to the terminal voltage, and the terminal-voltage magnitude. The source-port relationships depend on the synchronization and control dynamics.

With external references held fixed, the linearized SG and GFM dynamics relate power-injection perturbations to port-variable perturbations as $\Delta \mathbf{x}_i = M_i^{\mathrm{SG/GFM}}(s) \Delta \mathbf{p}_i$. For GFL resources, the PLL tracks the terminal-voltage phase, while converter controls regulate power injection, leading to $\Delta \mathbf{p}_i = -N_i^{\mathrm{GFL}}(s) \Delta \mathbf{x}_i$ [24], [25]. To accommodate these different input–output directions without transfer-matrix inversion, both relations are expressed in a common implicit form:

$$\boldsymbol{M}_i(s) \Delta \boldsymbol{p}_i + \boldsymbol{N}_i(s) \Delta \boldsymbol{x}_i = \mathbf{0} \tag{3}$$

where for SGs and GFMs, $(\boldsymbol{M}_i, \boldsymbol{N}_i) = \left(-\boldsymbol{M}_i^{\mathrm{SG/GFM}}, \boldsymbol{I}_2\right)$; for GFLs, $(\boldsymbol{M}_i, \boldsymbol{N}_i) = \left(\boldsymbol{I}_2, \boldsymbol{N}_i^{\mathrm{GFL}}\right)$, with $\boldsymbol{I}_2$ denoting the $2 \times 2$ identity matrix. This formulation allows different source types to be connected to a common network model while retaining their distinct synchronization and voltage-control dynamics.

The corresponding port matrices for representative SG, GFM, and DFIG-based GFL models are

$$\mathbf{M}^{\mathrm{SG}}(s) = \frac{\kappa}{\mathcal{D}} \begin{bmatrix} -G_r \mathcal{D} - a_{22} b_{11} + a_{12} b_{21} & -a_{22} b_{12} + a_{12} b_{22} \\ a_{11} b_{21} - a_{21} b_{11} & a_{11} b_{22} - a_{21} b_{12} \end{bmatrix} \tag{4}$$

$$\mathbf{M}^{\mathrm{GFM}}(s) = -F_p(s) \begin{bmatrix} \frac{\kappa\omega_0}{s\left(2Hs + D_{\mathrm{vsg}}^{\mathrm{pu}}\right)} & 0 \\ 0 & \frac{S_b}{V_{t0}}\left(n_q + \frac{K_{i,Q}}{s}\right) \end{bmatrix} \tag{5}$$

$$\mathbf{N}^{\mathrm{GFL}}(s) = \frac{1}{\kappa} \begin{bmatrix} \frac{K_{P\delta}(1 - G_{\mathrm{PLL}})}{D_P} & -\frac{K_{Pv}}{D_P} \\ \frac{K_{Q\delta}(1 - G_{\mathrm{PLL}})}{D_Q} & -\frac{K_{Qv}}{D_Q} \end{bmatrix} \tag{6}$$

where $\kappa = S_b / S_r$ converts power perturbations from the system base to the device base, $a_{mn}$ and $b_{mn}$ are the entries of $\boldsymbol{A}_{\mathrm{SG}}$ and $\boldsymbol{B}_{\mathrm{SG}}$, respectively, and $\mathcal{D} = \det(\boldsymbol{A}_{\mathrm{SG}})$. The coefficient matrices, transfer functions, and control parameters are specified in the Appendix.

### B. Network Model

The network model relates source power injections to bus voltage magnitudes and phase angles. To account for voltage-dependent load demand, constant-impedance loads are represented by equivalent shunt admittances and incorporated into the network admittance matrix. For a load at bus $k$,

$$Y_{L,k} = \frac{P_{L,k0} - jQ_{L,k0}}{V_{k0}^2}, \boldsymbol{Y}_N = \boldsymbol{Y}_{\mathrm{br}} + \mathrm{diag}\left(Y_{L,1}, \ldots, Y_{L,n_b}\right) \tag{7}$$

where $P_{L,k0}$ and $Q_{L,k0}$ denote positive load consumptions, $Y_{\mathrm{br}}$ is the branch-network admittance matrix, and $n_b$ is the total number of buses.

Let $G_{ij} + jB_{ij}$ denote element $(i,j)$ of $\boldsymbol{Y}_N$, and define $\theta_{ij} = \theta_i - \theta_j$. The nodal active and reactive power injections are

$$\begin{cases} P_i = \sum_{j=1}^{n_b} V_i V_j \left( G_{ij} \cos\theta_{ij} + B_{ij} \sin\theta_{ij} \right) \\ Q_i = \sum_{j=1}^{n_b} V_i V_j \left( G_{ij} \sin\theta_{ij} - B_{ij} \cos\theta_{ij} \right) \end{cases} \tag{8}$$

With $\Delta\boldsymbol{v} = \mathrm{diag}(\boldsymbol{V}_0)^{-1}\Delta\boldsymbol{V}$, linearizing the nodal power equations at $(\boldsymbol{\theta}_0, \boldsymbol{V}_0)$ yields

$$\begin{bmatrix} \Delta\boldsymbol{P} \\ \Delta\boldsymbol{Q} \end{bmatrix} = \underbrace{\begin{bmatrix} \mathbf{J}_{P\theta} & \mathbf{J}_{PV} \\ \mathbf{J}_{Q\theta} & \mathbf{J}_{QV} \end{bmatrix}}_{\mathbf{J}_N} \begin{bmatrix} \Delta\boldsymbol{\theta} \\ \Delta\boldsymbol{v} \end{bmatrix}, \begin{bmatrix} \Delta\boldsymbol{p}_g \\ \Delta\boldsymbol{p}_\ell \end{bmatrix} = \begin{bmatrix} \mathbf{J}_{gg} & \mathbf{J}_{g\ell} \\ \mathbf{J}_{\ell g} & \mathbf{J}_{\ell\ell} \end{bmatrix} \begin{bmatrix} \Delta\boldsymbol{x}_g \\ \Delta\boldsymbol{x}_\ell \end{bmatrix} \tag{9}$$

where the subscripts $g$ and $\ell$ identify variables at source and passive buses, respectively. The entries of $\mathbf{J}_N$ follow the power-flow Jacobian expressions in [23], with the voltage columns scaled by the corresponding operating-point magnitudes. For constant-impedance loads, $\Delta P_{L,k} = 2P_{L,k0}\Delta v_k$ and $\Delta Q_{L,k} = 2Q_{L,k0}\Delta v_k$. Thus, separating the load contribution gives

$$\boldsymbol{J}_N = \boldsymbol{J}_{\mathrm{br}} + 2\begin{bmatrix} \mathbf{0} & \mathit{diag}(\boldsymbol{P}_{L0}) \\ \mathbf{0} & \mathit{diag}(\boldsymbol{Q}_{L0}) \end{bmatrix} \tag{10}$$

where $\boldsymbol{J}_{\mathrm{br}}$ is obtained by linearizing (8) using $\boldsymbol{Y}_{\mathrm{br}}$ in place of $\boldsymbol{Y}_N$, and $\boldsymbol{P}_{L0}$ and $\boldsymbol{Q}_{L0}$ collect the operating-point load consumptions, with zeros at unloaded buses.

### C. Sparse Implicit Device-Network Closed Loop

An increase in active and reactive load demand at a selected passive bus is represented by $\Delta d = [\Delta P_L, \Delta Q_L]^T$. The matrix $E_d \in \mathbb{R}^{2n_\ell \times 2}$ places this disturbance at the selected bus, where $n_\ell$ is the number of passive buses. Since bus power is positive for injection, the corresponding passive-bus injection perturbation is $\Delta p_\ell = -E_d\Delta d$. For $n_g$ source devices, define $M(s) = \mathrm{blkdiag}\{M_i(s)\}$ and $N(s) = \mathrm{blkdiag}\{N_i(s)\}$. Combining (3) with the network equations in (9) yields

$$\begin{bmatrix} \boldsymbol{M}(s) & \boldsymbol{N}(s) & \mathbf{0} \\ \boldsymbol{I} & -\mathbf{J}_{gg} & -\mathbf{J}_{g\ell} \\ \mathbf{0} & \mathbf{J}_{\ell g} & \mathbf{J}_{\ell\ell} \end{bmatrix} \begin{bmatrix} \Delta\boldsymbol{p}_g \\ \Delta\boldsymbol{x}_g \\ \Delta\boldsymbol{x}_\ell \end{bmatrix} + \begin{bmatrix} \mathbf{0} \\ \mathbf{0} \\ \boldsymbol{E}_d \end{bmatrix} \Delta\boldsymbol{d} = \mathbf{0} \tag{11}$$

The first block row enforces device port constraints; the other two impose power balance at source and passive buses, completing the closed device–network feedback loop. The shared terminal variables couple the individual source dynamics through the network, so disturbances at passive buses can excite responses in all connected source devices.

The source-port relations are realized separately and assembled into the finite-dimensional form [26]

$$\begin{cases} \dot{\boldsymbol{\xi}} = \boldsymbol{A}_r\boldsymbol{\xi} + \boldsymbol{B}_{rp}\Delta\boldsymbol{p}_g + \boldsymbol{B}_{rx}\Delta\boldsymbol{x}_g \\ \mathbf{0} = \boldsymbol{C}_r\boldsymbol{\xi} + \boldsymbol{D}_{rp}\Delta\boldsymbol{p}_g + \boldsymbol{D}_{rx}\Delta\boldsymbol{x}_g \end{cases} \tag{12}$$

where $\xi \in \mathbb{R}^{n_r}$ contains the realization states. The matrices $A_r$, $B_{rp}$, and $B_{rx}$ define the state dynamics, while $C_r$, $D_{rp}$, and $D_{rx}$ define the port constraint. Specifically, under zero initial conditions, they satisfy

$$[\boldsymbol{M}(\boldsymbol{s})\ \boldsymbol{N}(\boldsymbol{s})] = [\boldsymbol{D}_{rp}\ \boldsymbol{D}_{rx}] + \boldsymbol{C}_r(s\boldsymbol{I} - \boldsymbol{A}_r)^{-1}[\boldsymbol{B}_{rp}\ \boldsymbol{B}_{rx}]. \tag{13}$$

Replacing the source-port constraint in (11) with (12) and defining $\boldsymbol{z} = \mathrm{col}\{\xi, \Delta p_g, \Delta x_g, \Delta x_\ell\}$ yields

$$\boldsymbol{E}\dot{\boldsymbol{z}} = \boldsymbol{A}\boldsymbol{z} + \boldsymbol{B}\boldsymbol{\Delta d} \tag{14}$$

where

$$\begin{cases} \boldsymbol{E} = \mathrm{blkdiag}\{\boldsymbol{I}_{n_r}, \mathbf{0}_{2n_g}, \mathbf{0}_{2n_g}, \mathbf{0}_{2n_\ell}\} \\ \boldsymbol{A} = \begin{bmatrix} \boldsymbol{A}_r & \boldsymbol{B}_{rp} & \boldsymbol{B}_{rx} & \mathbf{0} \\ \boldsymbol{C}_r & \boldsymbol{D}_{rp} & \boldsymbol{D}_{rx} & \mathbf{0} \\ \mathbf{0} & \boldsymbol{I}_{2n_g} & -\boldsymbol{J}_{gg} & -\boldsymbol{J}_{g\ell} \\ \mathbf{0} & \mathbf{0} & \boldsymbol{J}_{\ell g} & \boldsymbol{J}_{\ell\ell} \end{bmatrix}, \boldsymbol{B} = \begin{bmatrix} \mathbf{0} \\ \mathbf{0} \\ \mathbf{0} \\ \boldsymbol{E}_d \end{bmatrix}. \end{cases} \tag{15}$$

The passive-bus voltages remain algebraic variables. Thus, the descriptor model retains the local source-port blocks and the sparsity of the full-bus Jacobian without explicitly inverting the port operators or eliminating passive buses. Fig. 2 summarizes the construction of the sparse implicit device–network closed-loop model from the individual source-port models and network equations.

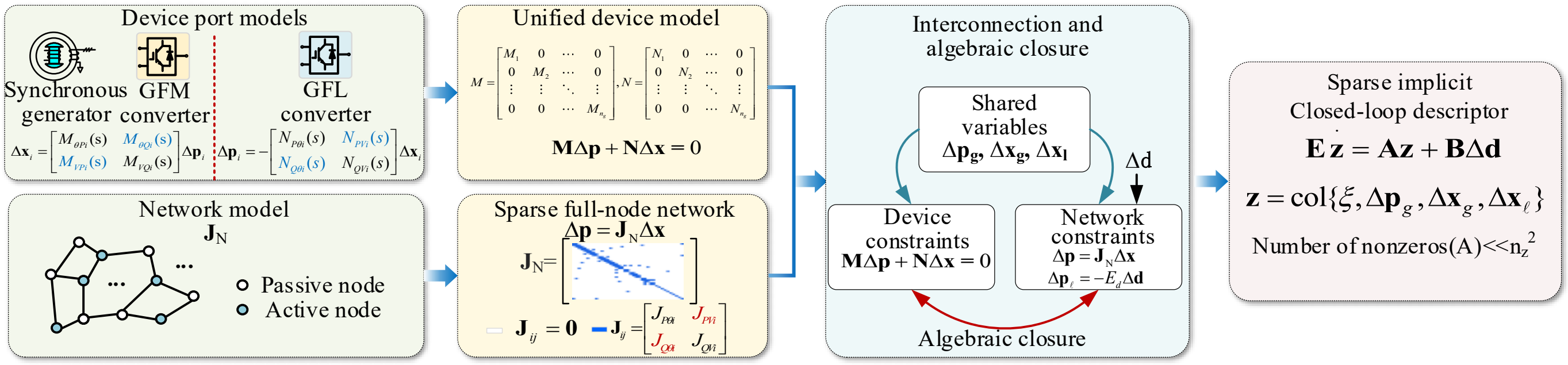

**Fig. 2.** Modeling framework for the sparse implicit source-network closed loop.

## III. Nodal Frequency Decomposition and Coupling Quantification

In this section, nodal frequency responses from the model in Section II are first decomposed into common- and differential-mode components. Then, closed-loop poles and residues characterize differential-mode dynamics, while frequency indices quantify spatial nonuniformity and frequency-voltage coupling effects.

### A. Nodal Frequency Decomposition and Modal Analysis

For a system with $n$ source devices, the closed-loop model in (14) yields the frequency response $\Delta\boldsymbol{f}(t) = [\Delta f_1(t), \dots, \Delta f_n(t)]^T$ to the load disturbance $\boldsymbol{\Delta d}$. Define the capacity weights $w_i = S_i / \sum_j S_j$ and $\boldsymbol{W} = \mathrm{diag}(S_i/S_b)$.

To separate aggregate frequency motion from relative variations among sources, a capacity-weighted spatial basis $\boldsymbol{\Phi} = [\boldsymbol{\phi}_1, \dots, \boldsymbol{\phi}_n]$ is constructed from the network coupling characteristics. The basis vectors are obtained by solving the following generalized eigenvalue problem and are normalized with respect to $\boldsymbol{W}$

$$L_\theta \phi_r = \rho_r \boldsymbol{W} \phi_r,\ \boldsymbol{\Phi}^T \boldsymbol{W} \boldsymbol{\Phi} = \boldsymbol{I}_n \tag{16}$$

where $L_\theta$ is obtained by eliminating passive-bus variables from the network Jacobian in (9) and symmetrizing the resulting source-angle coupling matrix for spatial basis construction. Its diagonal entries are then adjusted to ensure zero row sums. The zero row sums ensure that the uniform vector is an eigenvector representing common frequency motion. The remaining basis vectors describe relative frequency variations among sources.

The spatial coordinates are $\boldsymbol{\eta}(t) = \Phi^T W \Delta \mathbf{f}(t)$. With $\phi_1 = \mathbf{1}_n / \sqrt{\mathbf{1}_n^T W \mathbf{1}_n}$, the response is reconstructed as

$$\Delta f(t) = \mathbf{1}_n \Delta f_{\mathrm{CM}}(t) + \underbrace{\textstyle\sum_{r=2}^{n} \phi_r \eta_r(t)}_{\Delta f_{\mathrm{DM}}(t)} \tag{17}$$

where $\mathbf{1}_n$ and $I_n$ are the all-ones vector and identity matrix, respectively. The common-mode frequency is $\Delta f_{\mathrm{CM}}(t) = \boldsymbol{w}^T \Delta \boldsymbol{f}(t)$. Accordingly, the differential-mode projection is $\boldsymbol{P}_{\mathrm{DM}} = \boldsymbol{I}_n - \mathbf{1}_n \boldsymbol{w}^T$, which gives $\Delta \boldsymbol{f}_{\mathrm{DM}} = \boldsymbol{P}_{\mathrm{DM}} \Delta \boldsymbol{f}$.

To further characterize the temporal evolution of these components, the closed-loop dynamics in (14) are analyzed through their poles and residues. Under zero initial conditions, the disturbance-to-frequency transfer matrix $G_f(s)$ can be expanded at its simple finite poles as

$$\Delta \boldsymbol{f}(s) = \boldsymbol{G}_f(s) \Delta \boldsymbol{d}(s), \boldsymbol{G}_f(s) = \boldsymbol{D}_f + \textstyle\sum_k \frac{\boldsymbol{R}_k}{s - \lambda_k} \tag{18}$$

where $\lambda_k$ is the $k$th pole of $\boldsymbol{G}_f(s)$, $\boldsymbol{R}_k = \mathrm{Res}_{s=\lambda_k} \boldsymbol{G}_f(s)$ is its residue matrix, and $\boldsymbol{D}_f$ is the direct-feedthrough matrix.

For a step disturbance $\Delta \mathbf{d}(t) = \mathbf{d}_0 u(t)$, projecting the frequency response onto the differential-mode subspace gives

$$\Delta \boldsymbol{f}_{\mathrm{DM}}(t) = \boldsymbol{P}_{\mathrm{DM}} \boldsymbol{D}_f \boldsymbol{d}_0 + \textstyle\sum_k \boldsymbol{a}_k h_k(t), \tag{19}$$

$$\boldsymbol{a}_k = \boldsymbol{P}_{\mathrm{DM}} \boldsymbol{R}_k \boldsymbol{d}_0, h_k(t) = \begin{cases} (e^{\lambda_k t} - 1)/\lambda_k, & \lambda_k \neq 0 \\ t, & \lambda_k = 0 \end{cases} \tag{20}$$

where $u(t)$ is the unit-step function and $\boldsymbol{d}_0$ is the disturbance magnitude vector. Each pole $\lambda_k$ determines the oscillation frequency and decay rate of its contribution, while $\boldsymbol{a}_k$ describes its spatial amplitude and phase under the specified disturbance.

The spatial basis separates common and relative frequency motion, whereas the pole–residue expansion resolves their temporal dynamics. A single spatial component may contain contributions from multiple closed-loop poles; conversely, one pole may contribute to several spatial components through its residue matrix.

Thus, (19) and (20) connect the device–network dynamics retained in Section II to the spatial frequency differences quantified in Section III-B.

### *B. Quantification of Frequency-Voltage Coupling*

To quantify spatial frequency nonuniformity among source buses, define the instantaneous spatial standard deviation as

$$D_{\mathrm{sp}}(t) = \frac{\left\| \boldsymbol{P}_{\mathrm{sp}} \Delta \boldsymbol{f}(t) \right\|_2}{\sqrt{n}} \tag{21}$$

where $\boldsymbol{P}_{\mathrm{sp}} = \boldsymbol{I}_n - \frac{1}{n} \mathbf{1}_n \mathbf{1}_n^T$ removes the mean frequency across source buses. Thus, $D_{\mathrm{sp}}(t)$ measures the instantaneous frequency differences.

To account for both the magnitude and duration of nodal frequency differences, the root-mean-square value of $D_{\mathrm{sp}}(t)$ over $[t_a, t_b]$ is defined as

$$D_{\mathrm{ST}} = \sqrt{\frac{1}{T} \int_{t_a}^{t_b} D_{\mathrm{sp}}^2(t) dt},\ T = t_b - t_a \tag{22}$$

For a fixed analysis window, a larger $D_{\mathrm{ST}}$ indicates greater or more persistent spatial frequency nonuniformity.

To assess the effect of F-V coupling, the full and decoupled models are evaluated at the same operating point under the same disturbance. The full model retains the device-port and network coupling channels established in Section II, whereas the decoupled model retains only the $P - \theta$ and $Q - V$ channels. Their difference in time-aggregated spatial nonuniformity is

$$\Delta D_{\text{FV-ST}} = D_{\mathrm{ST}}^{\mathrm{Full}} - D_{\mathrm{ST}}^{P-\theta} \tag{23}$$

where the superscript $P - \theta$ denotes the decoupled model.

A positive value indicates that F-V coupling increases spatial frequency nonuniformity; a negative value indicates the opposite. The indices are interpreted alongside closed-loop poles and frequency responses in Section IV to explain the effects of F-V coupling.

## IV. Effects of Frequency-Voltage Coupling on Nodal Frequency

This section applies the decomposition and coupling indices in Section III to examine how F-V coupling, source configuration, and control and network parameters affect nodal frequency differences. The analysis considers the modified 10-machine, 39-bus system shown in Fig. 3. Source configurations and disturbances are summarized in Table I, with device parameters provided in the Appendix.

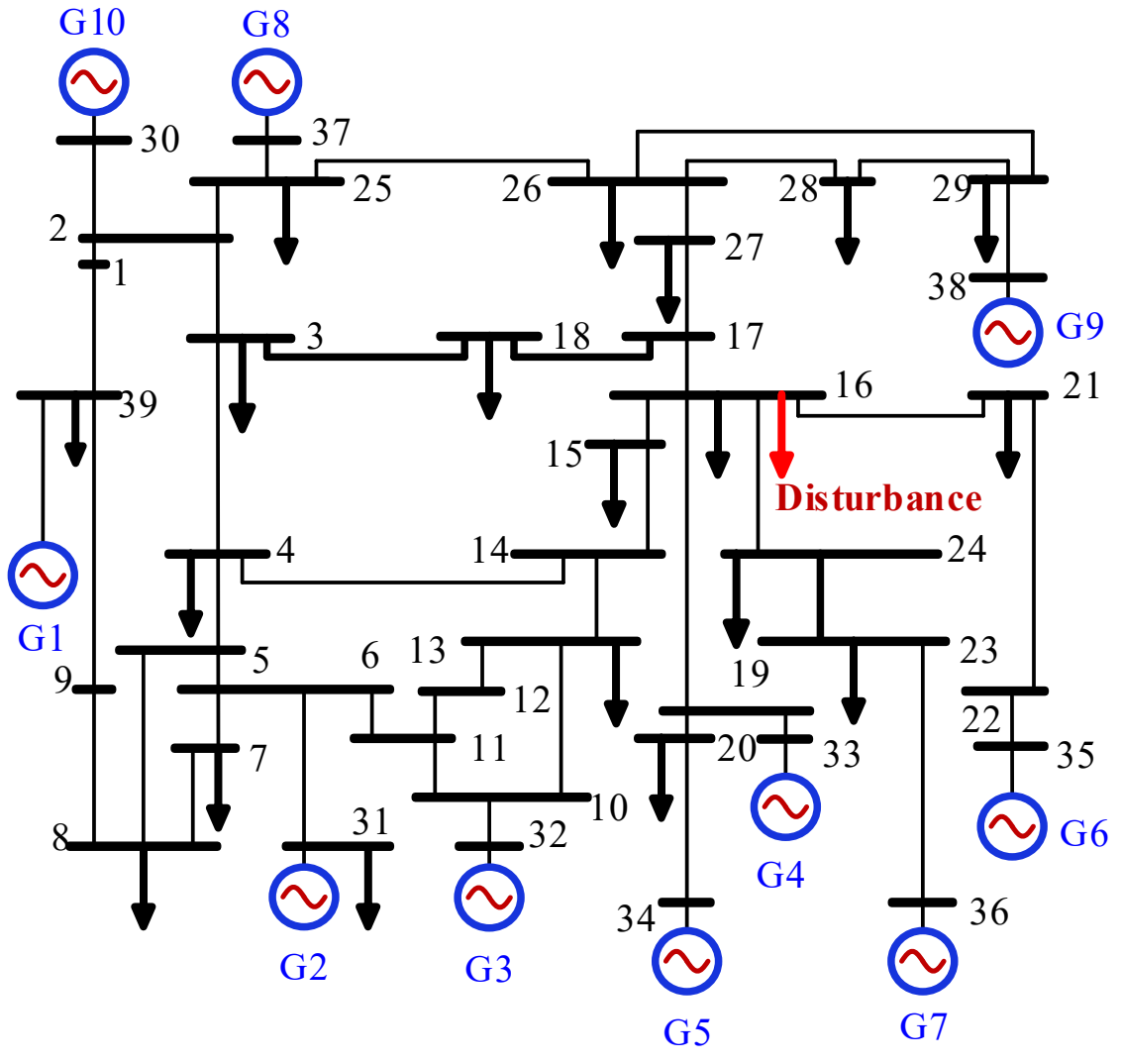

**Fig. 3.** Modified 10-machine, 39-bus test system.

TABLE I
SOURCE CONFIGURATIONS AND DISTURBANCES

| Case | Generation configuration | Disturbance Type |
|---|---|---|
| 1-a | G5, G7, G8: DFIG(GFL); G3, G4, G10: GFM; the rest: SG. | Bus 16, $\Delta P_L$=500 MW |
| 1-b | Same as Case 1-a. | Bus 16, $\Delta Q_L$=500MVar |
| 2-a | All units: SG. | Bus 16, $\Delta P_L$=500 MW |
| 2-b | G5, G7, G8: DFIG(GFL); the rest: SG. | Bus 16, $\Delta P_L$=500 MW |
| 2-c | G3, G4, G5, G7, G8, G10: DFIG(GFL); the rest: SG. | Bus 16, $\Delta P_L$=500 MW |
| 2-d | G3, G4, G10: GFM; the rest: SG. | Bus 16, $\Delta P_L$=500 MW |
| 2-e | G3, G4, G5, G7, G8, G10: GFM; the rest: SG. | Bus 16, $\Delta P_L$=500 MW |
| 3-a | Same as Case 1-a. | Bus 8, $\Delta P_L$=500 MW |
| 3-b | Same as Case 1-b. | Bus 16, $\Delta P_L$=300 MW |

### *A. Spatial Frequency Response*

For Case 1-a, the full and decoupled models are compared using the dispersion indices and pole analysis in Section III to assess F-V coupling effects.

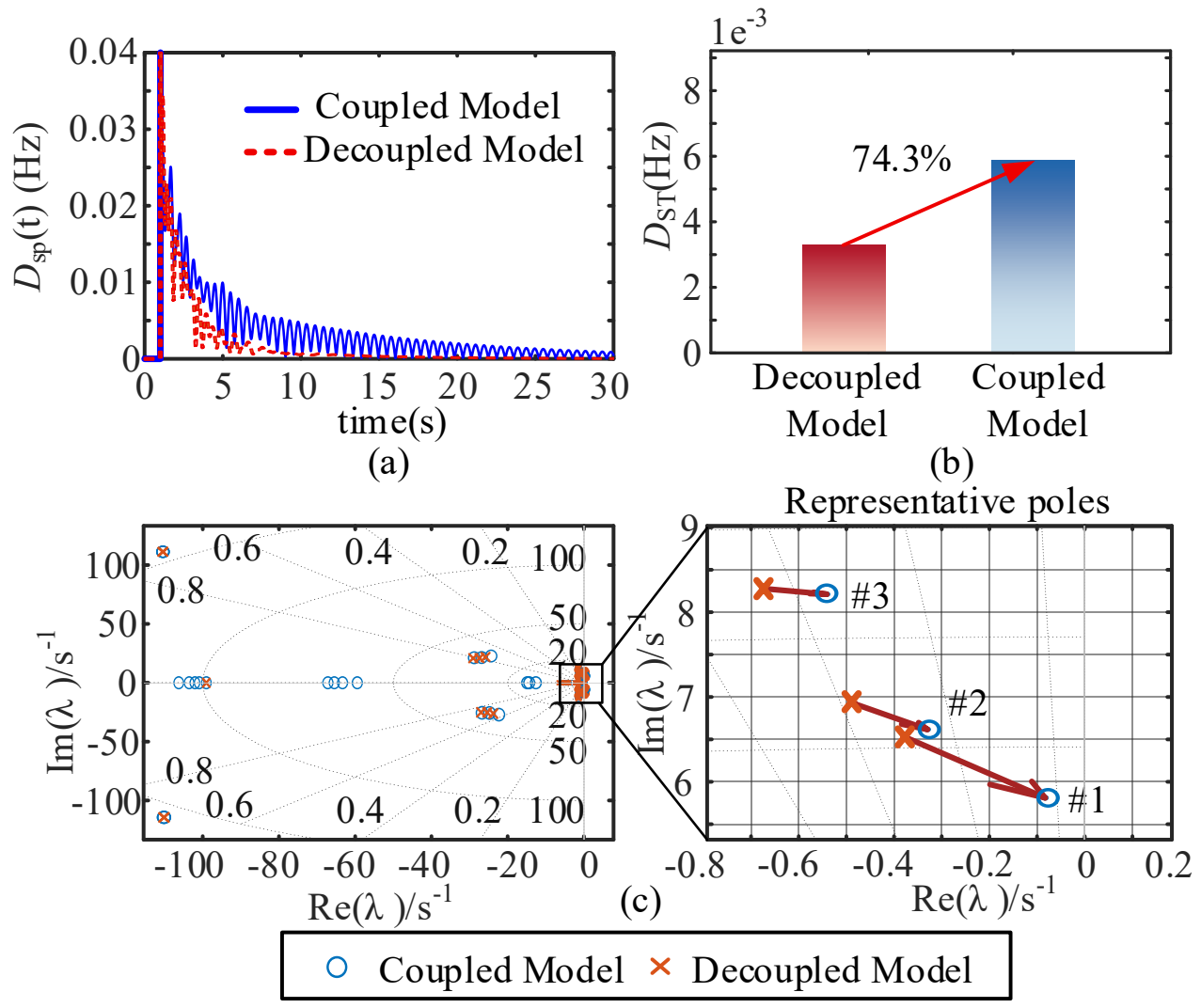

**Fig. 4.** F-V coupling effects in Case 1-a. (a) $D_{sp}(t)$. (b) $D_{ST}$. (c) Representative closed-loop poles.

Fig. 4(a) shows larger and more persistent spatial frequency dispersion with F-V coupling, while Fig. 4(b) shows a 74.3% increase in $D_{ST}$. Under the decomposition in (17), this dispersion reflects relative frequency variations captured by the differential component. The representative poles in Fig. 4(c) move toward the imaginary axis, indicating slower decay of the transient terms $e^{\lambda_k t}$ in (20). This is consistent with the prolonged differential-mode response. The contribution of each pole to spatial dispersion also depends on the projected residue coefficient $\boldsymbol{a}_k$ in (20).

### *B. Effects of Source Configuration*

Starting from the all-SG baseline in Case 2-a, SGs are progressively replaced by DFIG-based GFL or GFM units in Cases 2-b–2-e and 1-a. These replacements modify the source-port matrices in (3), thereby changing the closed-loop frequency transfer matrix in (18).

The spatial dispersion is evaluated using $D_{ST}$ in (22). To identify the frequency bands affected by F-V coupling, define

$$\Gamma_{FV}(f) = \left\| \boldsymbol{P}_{sp}\left[\boldsymbol{G}_f^{Full}(j2\pi f) - \boldsymbol{G}_f^{P-\theta}(j2\pi f)\right]\boldsymbol{e}_P \right\|_2 \quad (24)$$

where $\mathbf{e}_P = [1,0]^T$ selects a unit active-load disturbance, and $\mathbf{G}_f^{Full}$ and $\mathbf{G}_f^{P-\theta}$ denote the transfer matrices relating load disturbances to nodal frequency responses in the full and decoupled models, respectively. $D_{ST}$ summarizes dispersion over time, and $\Gamma_{FV}$ identifies the frequency ranges in which coupling most strongly modifies the spatial response.

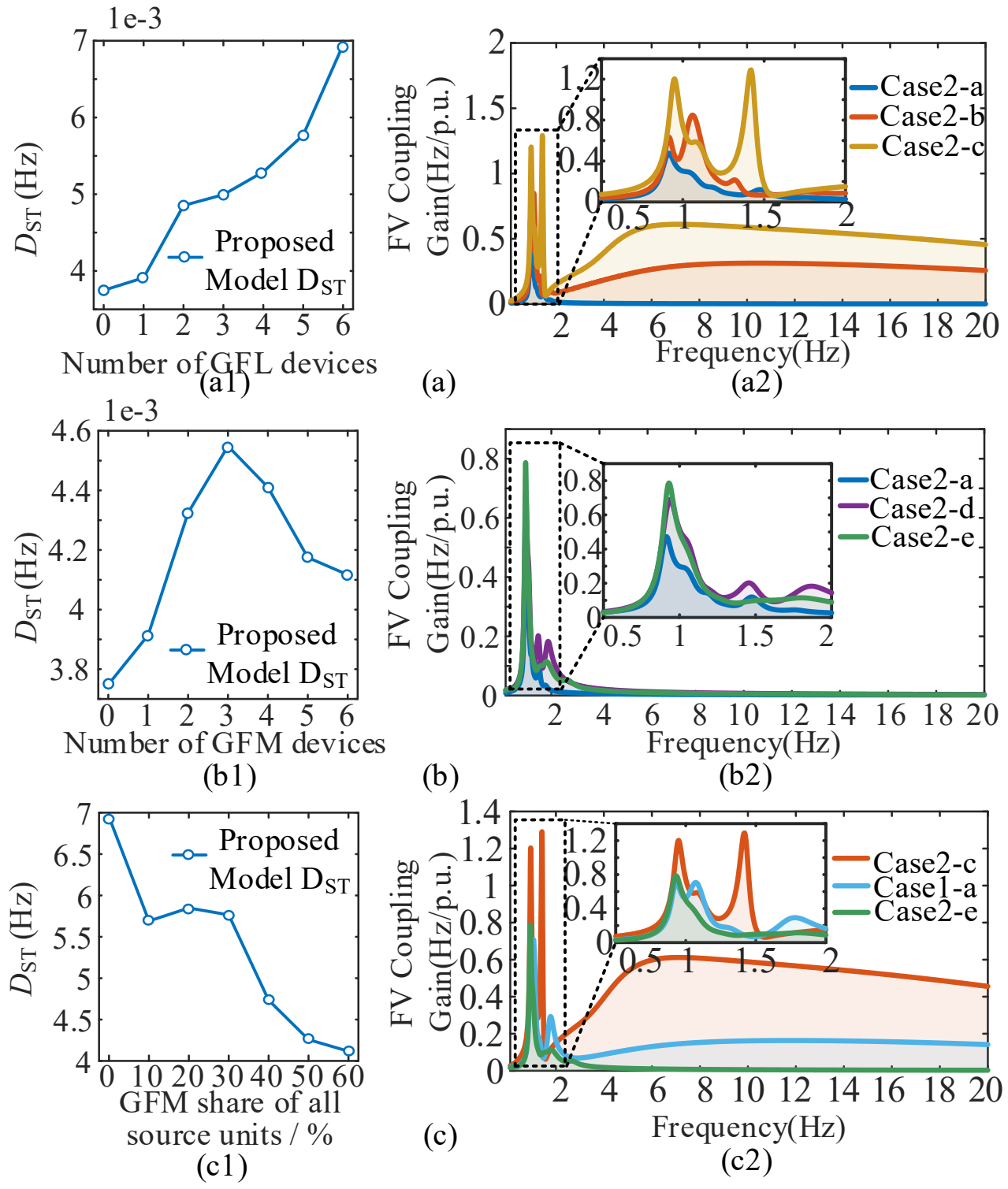

**Fig. 5.** Effects of source configuration on $D_{ST}$ (left) and $\Gamma_{FV}$ (right). (a) Increasing GFL count. (b) Increasing GFM count. (c) Increasing GFM share at a fixed converter count.

Fig. 5(a) shows that GFL replacement increases $D_{ST}$ and broadens the coupling response. Fig. 5(b) shows nonmonotonic dispersion under GFM replacement, with coupling concentrated at lower frequencies. In Fig. 5(c), increasing the GFM share at a fixed converter count reduces dispersion and higher-frequency coupling. These contrasting trends show that converter penetration alone does not determine spatial frequency nonuniformity; the synchronization and control dynamics represented in (3), together with their network interactions, also shape the response.

### *C. Parameter Sensitivity*

Case 1-a is retained to examine the control and network parameters underlying these configuration-dependent responses. Device parameters modify the source-port dynamics, while network impedance modifies the network Jacobian in Section II. The investigated parameters cover DFIG synchronization and power controls [18], GFM frequency and voltage controls [27], SG excitation, and line impedance [28].

The Morris elementary-effects method [29] screens their influence on the coupling-induced dispersion change $Y(\boldsymbol{x}) = \Delta D_{\text{FV-ST}}$, defined in (23). Over normalized parameter ranges, the sensitivity of parameter $i$ is measured by

$$\mu_i^* = \frac{1}{N_i}\sum_{r=1}^{N_i}\left|\frac{Y(\boldsymbol{x}^{(r)}+\delta \boldsymbol{e}_i)-Y(\boldsymbol{x}^{(r)})}{\delta}\right| \tag{25}$$

where $\mathbf{x}^{(r)}$ is a normalized parameter sample, $\mathbf{e}_i$ selects parameter $i$, $\delta$ is the normalized perturbation step, and $N_i$ is the number of valid elementary effects.

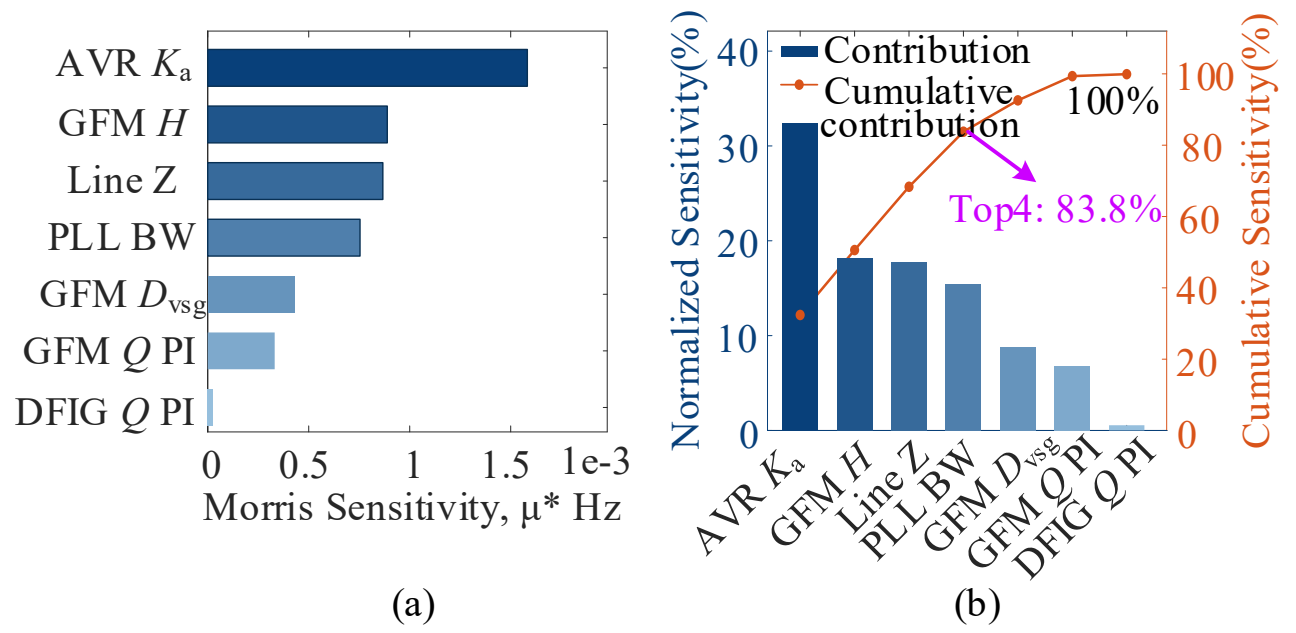

**Fig. 6.** Morris sensitivity analysis of $\Delta D_{\text{FV-ST}}$ in Case 1-a. (a) Parameter ranking. (b) Individual and cumulative contributions to $\sum_i \mu_i^*$.

Fig. 6 identifies SG AVR gain, GFM virtual inertia, the line-impedance multiplier, and DFIG PLL bandwidth as the four most influential parameters. The cumulative contribution curve shows an elbow after the fourth parameter, with these four accounting for 83.8% of the summed $\mu_i^*$. They are therefore selected for individual sweeps.

The effect of each parameter $p$ on full-model dispersion is measured relative to its nominal value $p_0$ by $R_D(p) = (D_{ST}^{full}(p) - D_{ST}^{full}(p_0))/D_{ST}^{full}(p_0) \times 100\%$

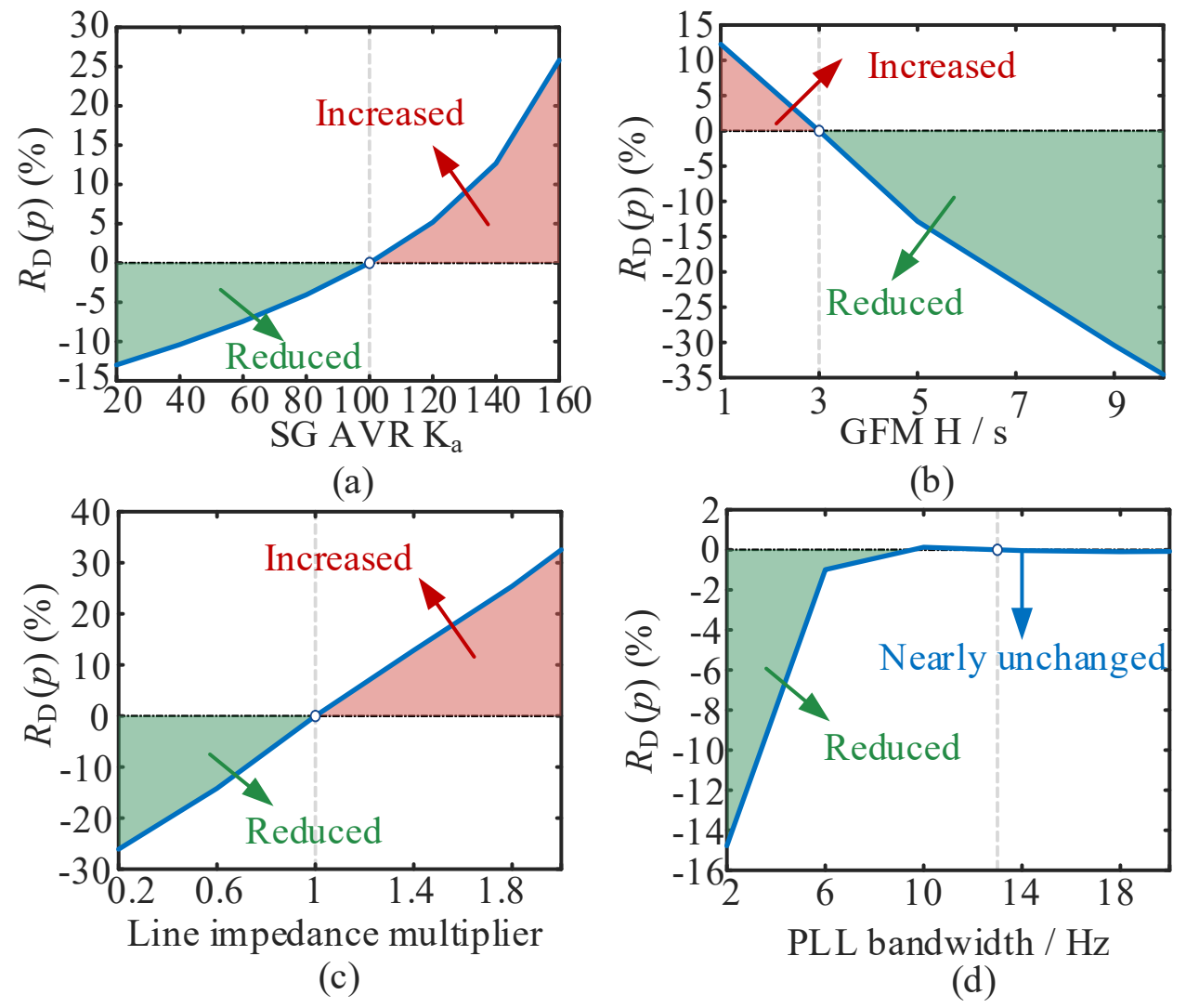

**Fig. 7.** Single-parameter sweeps of $R_D$. (a) SG AVR gain. (b) GFM virtual inertia. (c) Line-impedance multiplier. (d) DFIG PLL bandwidth.

Fig. 7 shows that increasing SG AVR gain or the line-impedance multiplier raises $D_{\text{ST}}$, whereas increasing GFM virtual inertia lowers it. The response to PLL bandwidth is strongest at low bandwidths and levels off as the bandwidth increases. These sweeps characterize how the parameters selected by Morris screening affect spatial frequency dispersion in the full model.

## V. Case Studies

In this section, the accuracy of the model developed in Section II and the source-configuration and parameter trends identified in Section IV are validated. The topology and cases follow Section IV, and computational efficiency is evaluated across systems of increasing size.

### *A. Accuracy of the Model and Frequency Decomposition*

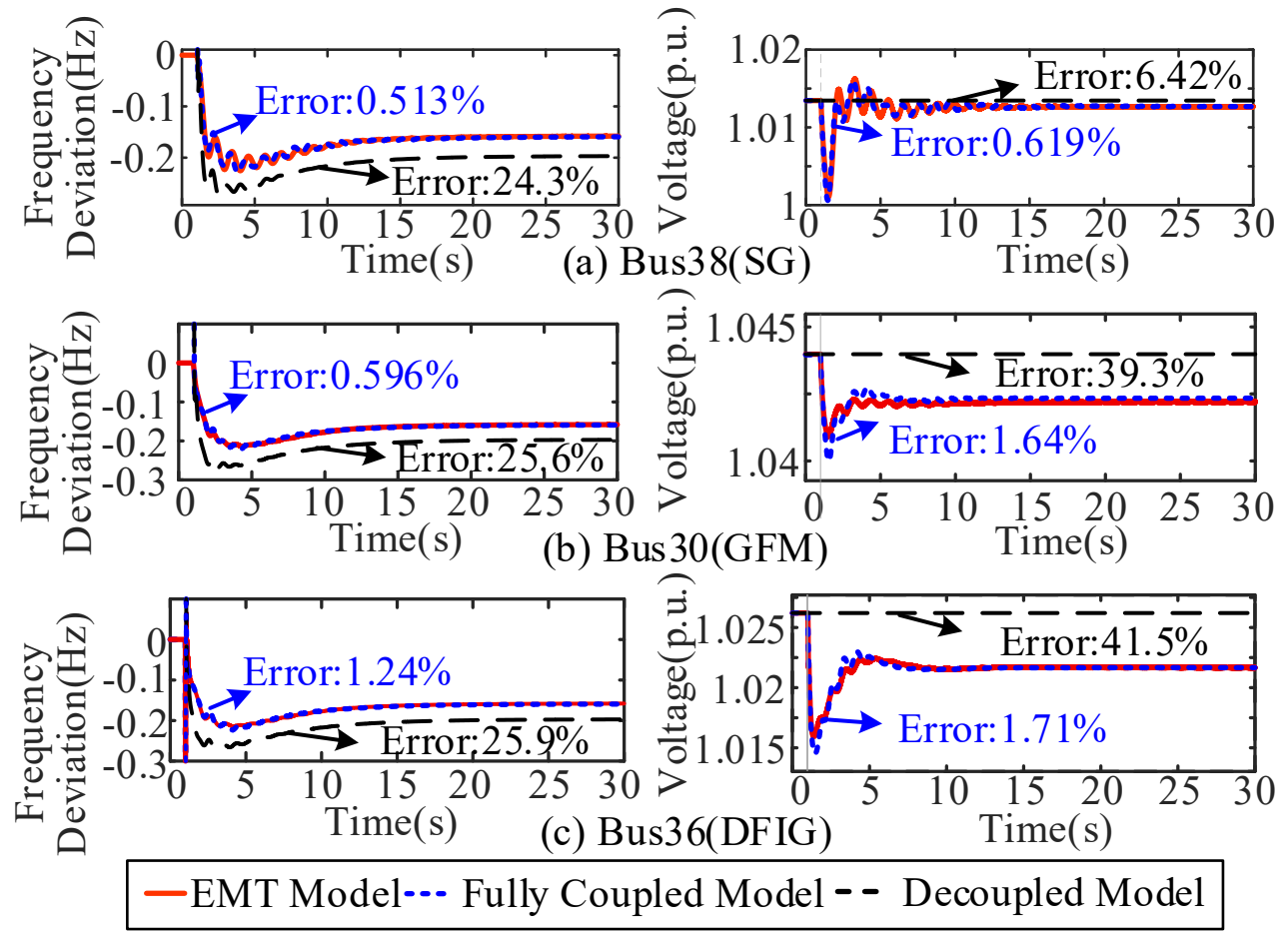

**Fig. 8.** Comparison of model predictions and EMT results in Case 1-a. (a) Bus 38 (SG). (b) Bus 30 (GFM). (c) Bus 36 (DFIG).

Fig. 8 compares the frequency and voltage responses predicted by the proposed and decoupled models with EMT simulations following a 500-MW active-power disturbance in Case 1-a. The three representative buses correspond to SG, GFM, and GFL resources. The proposed model closely reproduces both the transient and steady-state responses, with maximum relative errors of 1.24% for frequency and 1.71% for voltage. The decoupled model yields errors of up to 25.9% and 41.5%, respectively, demonstrating the importance of retaining F-V coupling for accurate response prediction.

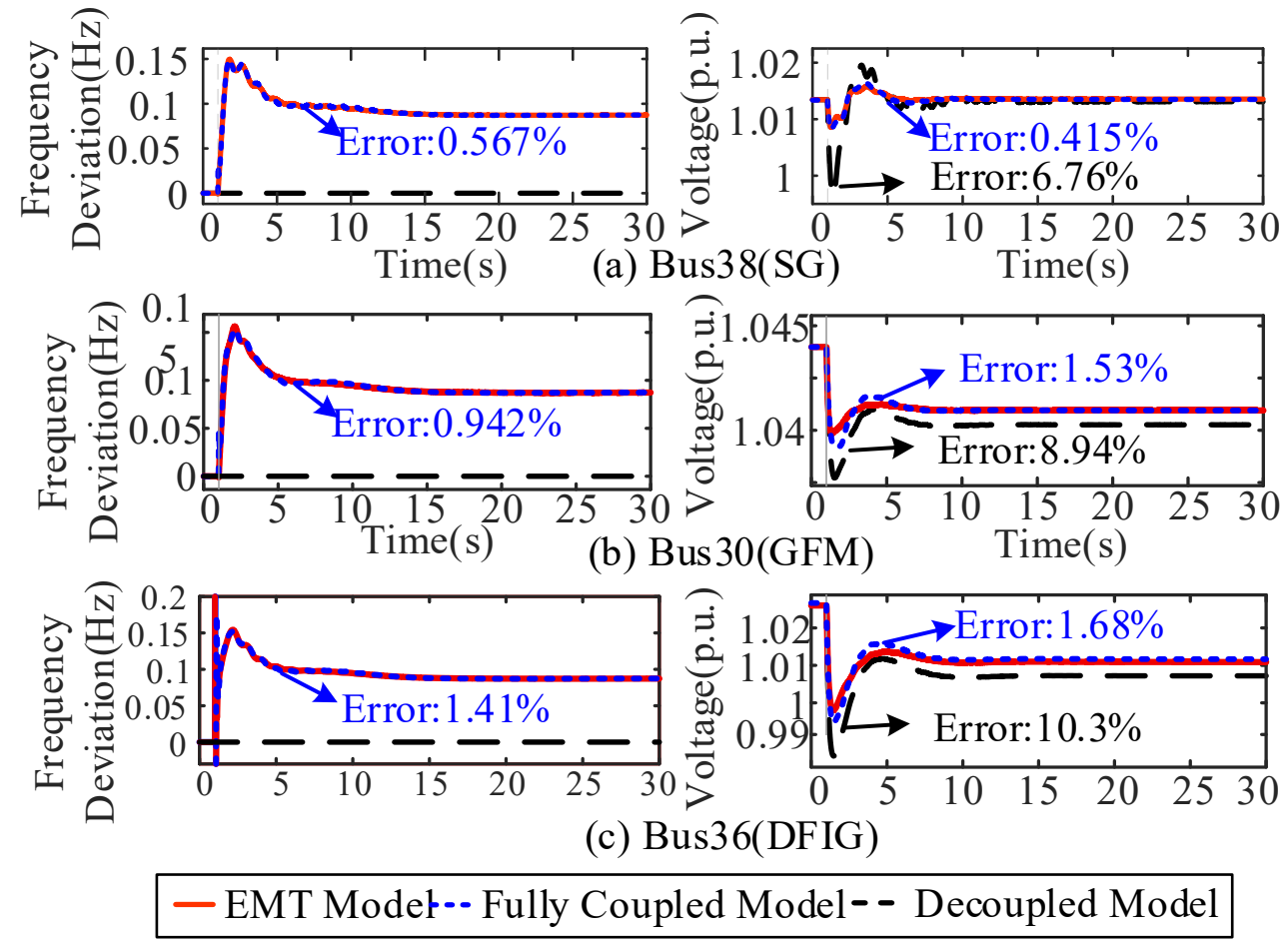

**Fig. 9.** Comparison of model predictions and EMT results in Case 1-b. (a) Bus 38 (SG). (b) Bus 30 (GFM). (c) Bus 36 (DFIG).

Fig. 9 extends the comparison to the 500-MVar reactive-power disturbance in Case 1-b. The proposed model captures the resulting frequency and voltage responses, with maximum relative errors of 1.41% and 1.68%, respectively. The decoupled model fails to reproduce the induced frequency response and exhibits voltage errors of up to 10.3%. This comparison demonstrates that reactive-power disturbances can also excite nodal frequency dynamics through F-V coupling, supporting the inclusion of both active- and reactive-power disturbance channels in the closed-loop model.

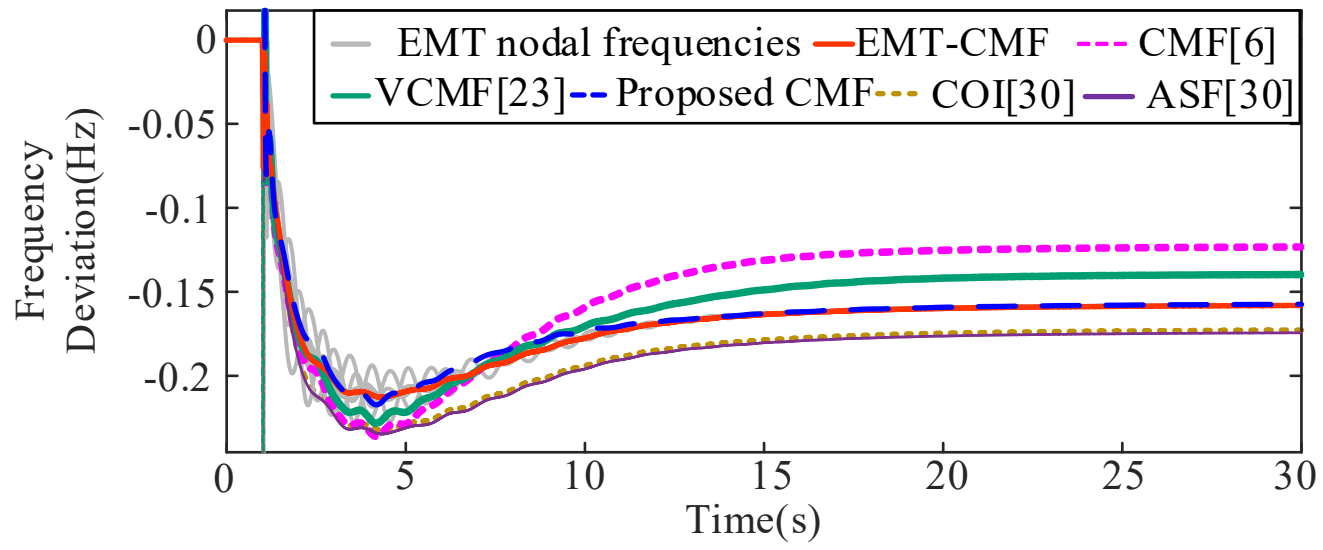

**Fig. 10.** Comparison of common-mode and aggregate frequency responses.

For Case 1-a, Fig. 10 compares the proposed common-mode frequency (CMF) with the EMT-derived CMF, the CMF in [6], the voltage-influenced CMF (VCMF) in [23], and the conventional ASF and COI frequencies [30]. The proposed CMF closely matches the EMT reference throughout the transient. This agreement extends to the steady state after the nodal oscillations subside. The CMF in [6] and VCMF predict deeper frequency nadirs but recover to higher frequencies than the EMT-derived CMF. ASF and COI also predict deeper nadirs, yet settle at lower frequencies than the reference.

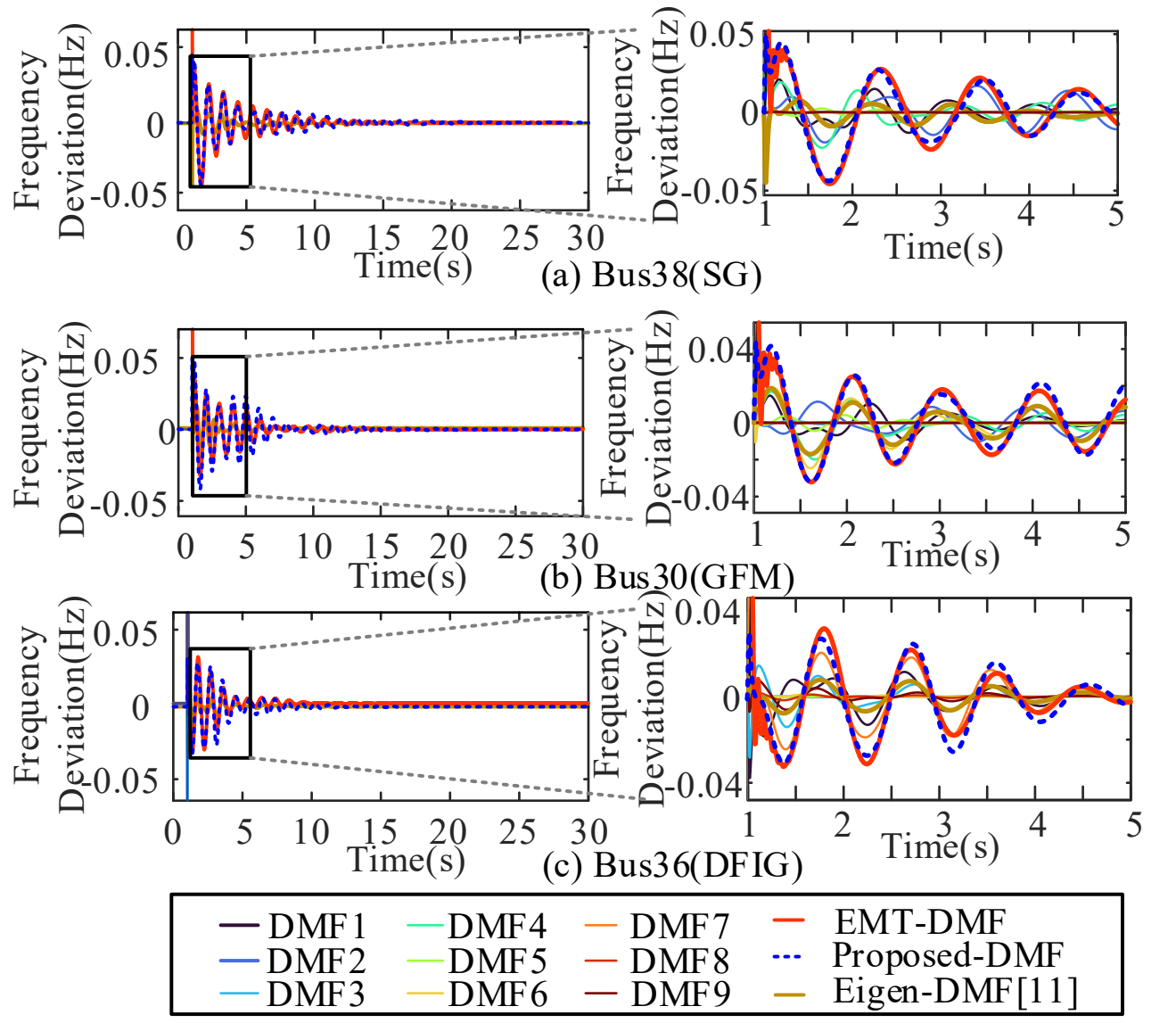

**Fig. 11.** Differential-mode frequency reconstruction. (a) Bus 38 (SG). (b) Bus 30 (GFM). (c) Bus 36 (DFIG).

Fig. 11 evaluates the differential-mode reconstruction in Case 1-a against EMT-derived responses and the characteristic-subsystem method [11]. The proposed method accurately reproduces the differential-mode oscillations at three representative buses, whereas the method in [11] exhibits larger deviations. These results demonstrate the accuracy of frequency reconstruction using the proposed model.

### *B. Validation of Source-Configuration and Parameter Effects*

To validate the source-configuration effects identified in Section IV-B, the frequency responses and $D_{ST}$ are compared with EMT results in Fig. 12. The comparisons cover GFL and GFM replacement and variations in their relative shares.

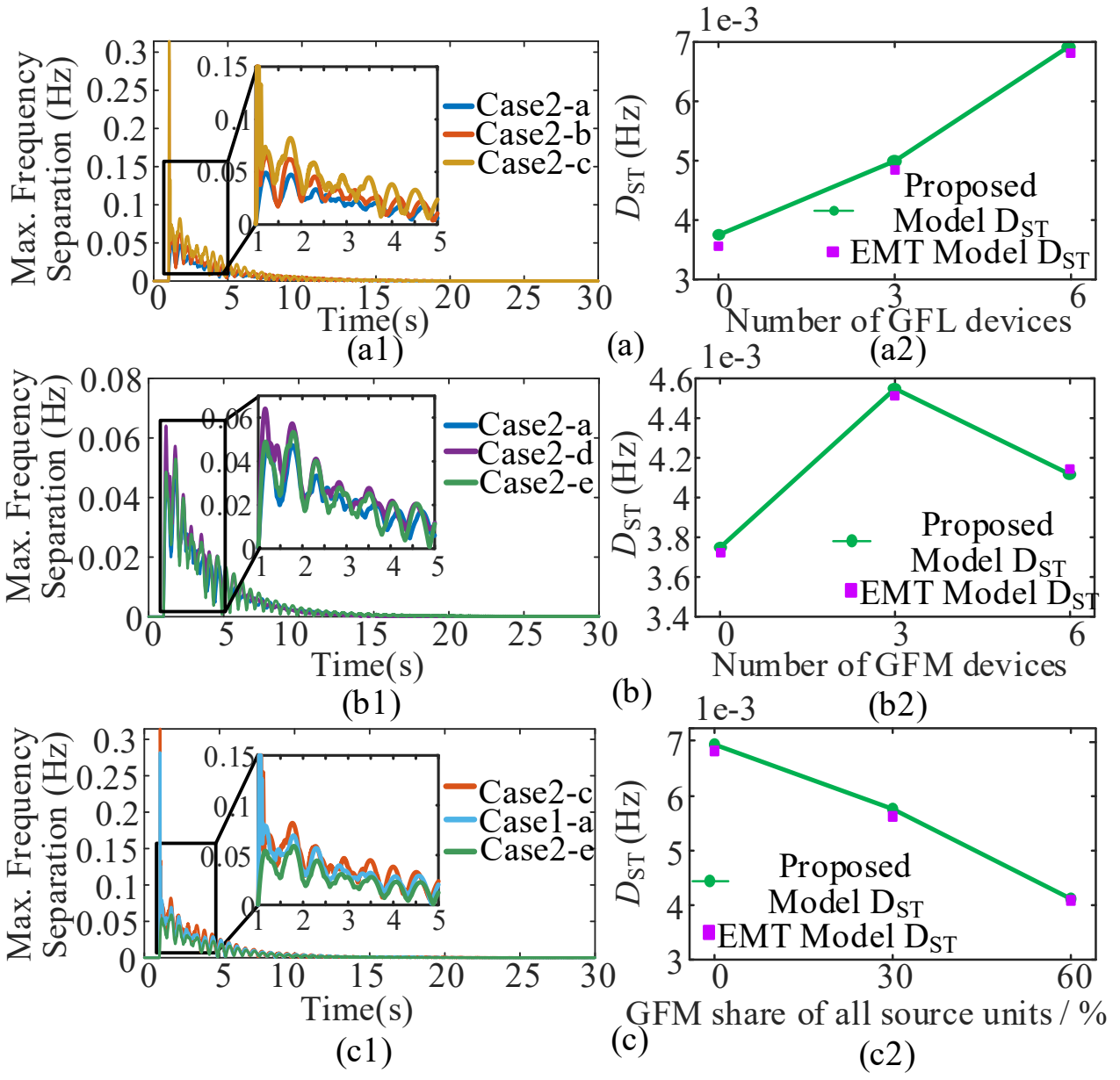

**Fig. 12.** Validation of source-configuration effects. (a) GFL replacement. (b) GFM replacement. (c) Varying GFL-GFM composition.

Fig. 12 shows that increasing GFL penetration raises $D_{ST}$ from $3.7\times10^{-3}$ to $6.9\times10^{-3}$ Hz, whereas GFM penetration produces a nonmonotonic trend. At a fixed converter count of six, increasing the GFM share reduces $D_{ST}$ from $6.9\times10^{-3}$ to $4.1\times10^{-3}$ Hz. The results agree with EMT simulations.

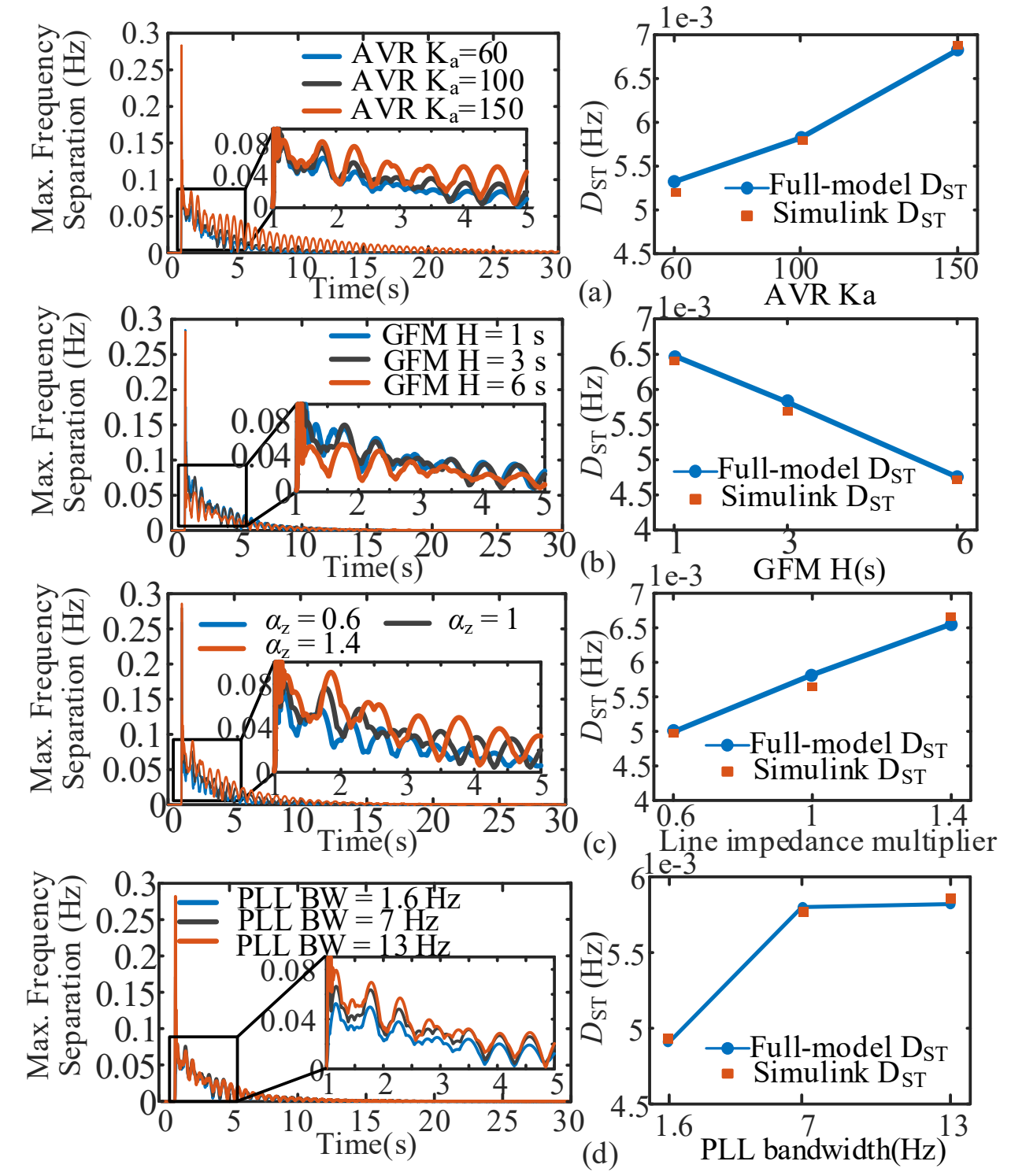

**Fig. 13.** Effects of key parameters on spatial frequency dispersion. (a) SG AVR gain. (b) GFM virtual inertia. (c) Line-impedance multiplier. (d) DFIG PLL bandwidth.

The parameter trends identified in Section IV-C are further evaluated against EMT simulations in Fig. 13. Increasing SG AVR gain and line impedance increases $D_{ST}$, whereas higher GFM virtual inertia reduces it. The influence of DFIG PLL bandwidth becomes limited above approximately 7 Hz. The model predictions closely match the EMT-based results, confirming the parameter-dependent trends.

### *C. Robustness and Computational Efficiency*

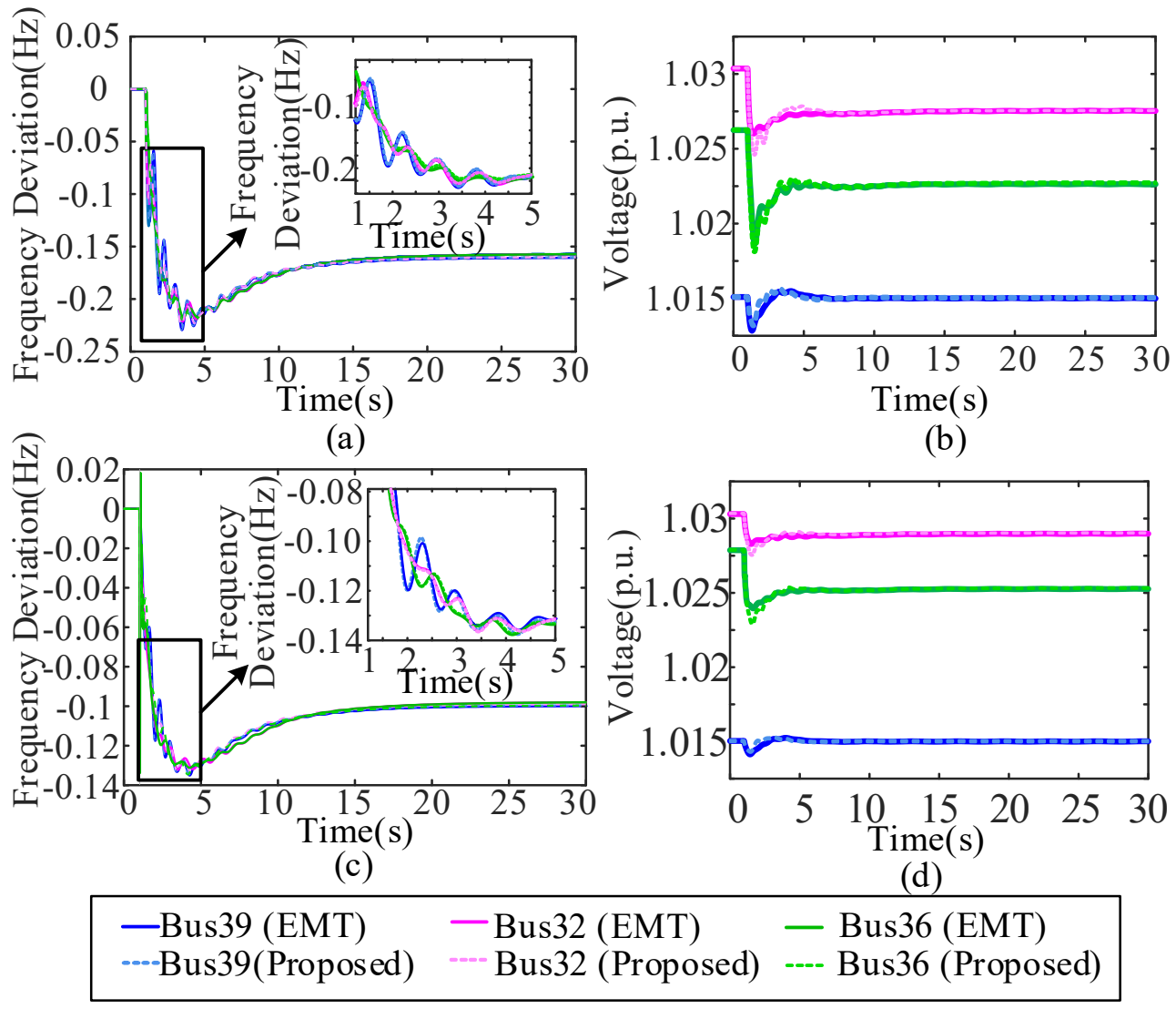

**Fig. 14.** Frequency and voltage responses in Cases 3-a and 3-b. (a) Frequency in Case 3-a. (b) Voltage in Case 3-a. (c) Frequency in Case 3-b. (d) Voltage in Case 3-b.

Fig. 14 compares the proposed model with EMT simulations in Cases 3-a and 3-b. The predicted frequency and voltage responses remain in close agreement with EMT results under changes in disturbance location and magnitude, demonstrating the model's robustness.

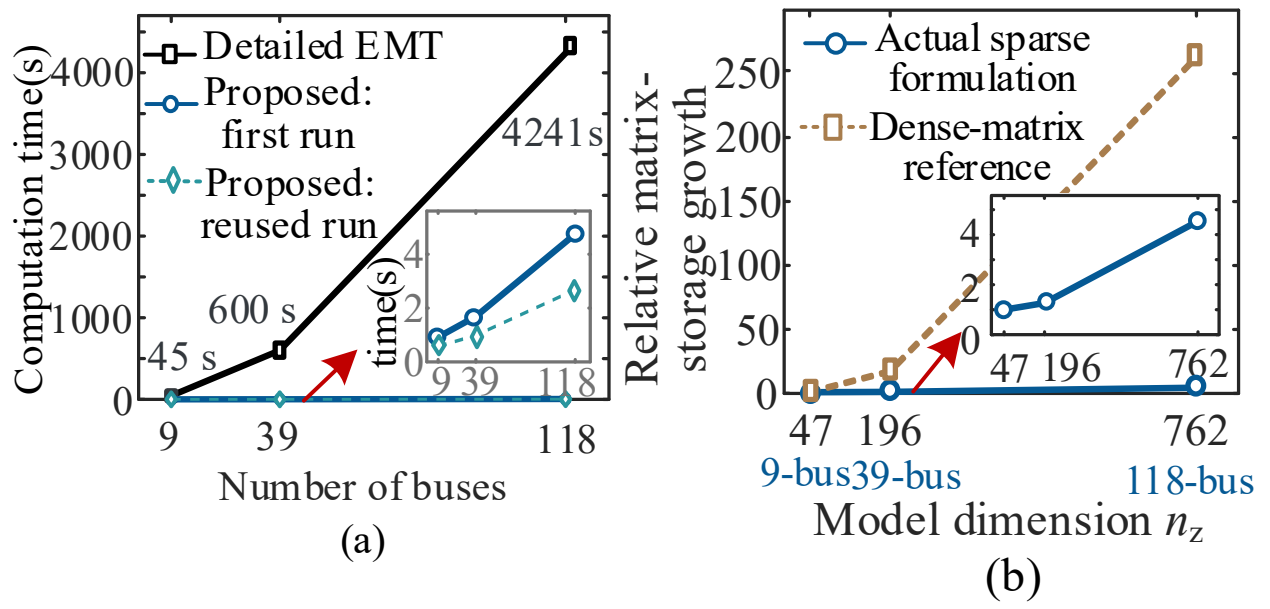

**Fig. 15.** Computational scalability of the sparse implicit model. (a) Computation time. (b) Matrix-storage growth.

TABLE II
COMPUTATIONAL PERFORMANCE AND MATRIX SPARSITY

| Metric | 9-bus | 39-bus | 118-bus |
|---|---|---|---|
| EMT simulation time (s) | 45 | 600 | 4241 |
| First-run time (s) | 0.860 | 1.576 | 4.638 |
| Reused-run time (s) | 0.556 | 0.990 | 2.714 |
| Matrix dimension $n_z$ | 47 | 196 | 762 |
| Nonzero entries nnz($\boldsymbol{A}$) | 933 | 1181 | 4233 |
| Matrix density (%) | 42.236 | 3.074 | 0.729 |

Fig. 15 and Table II compare computational performance across the 9-, 39-, and 118-bus systems. EMT simulation takes 45-4241 s, whereas the proposed model requires 0.860-4.638 s initially and 0.556-2.714 s with model reuse. The 118-bus descriptor matrix has a density of 0.729%, supporting efficient computation and sparse storage as system size increases.

## VI. CONCLUSION

This paper developed a sparse implicit device–network closed-loop model that retains F-V coupling among SG, GFL, and GFM resources. The analysis clarified how source configuration, control dynamics, and network parameters influence spatial differences in nodal frequency. The main conclusions are as follows.

1. **Retaining F-V coupling substantially improves frequency and voltage prediction accuracy.** Simulation errors remain within 1.41% and 1.71%, respectively, compared with up to 25.9% and 41.5% for the decoupled model. Accuracy is maintained under changes in disturbance type, location, and magnitude. These results demonstrate the need to retain voltage dynamics when assessing nodal frequency responses in systems integrating multiple source types.
2. **F-V coupling can amplify spatial frequency dispersion, with distinct effects under GFL and GFM synchronization.** The observed 74.3% increase in $D_{\mathrm{ST}}$ accompanies slower differential-mode decay. In the studied cases, dispersion increases with GFL penetration but varies nonmonotonically with GFM penetration. Replacing GFL with GFM reduces dispersion, showing that converter penetration alone cannot determine spatial frequency dispersion. Higher SG AVR gain and network impedance increase dispersion, whereas greater GFM inertia reduces it.
3. **The sparse implicit formulation substantially reduces computational cost and matrix storage growth.** For the 118-bus system, it achieves approximately 900-fold acceleration relative to EMT simulation, increasing to 1,560-fold with model reuse. Preserving network sparsity also limits matrix storage growth, supporting repeated disturbance analyses in larger systems.

## APPENDIX

This appendix specifies the device models and coefficients used in (4)-(6). The control diagrams and parameter tables describe the SG, GFM, and DFIG-based GFL implementations used in the case studies.

**1) Synchronous generator:** The SG structure and control parameters are shown in Fig. 16 and Table III.

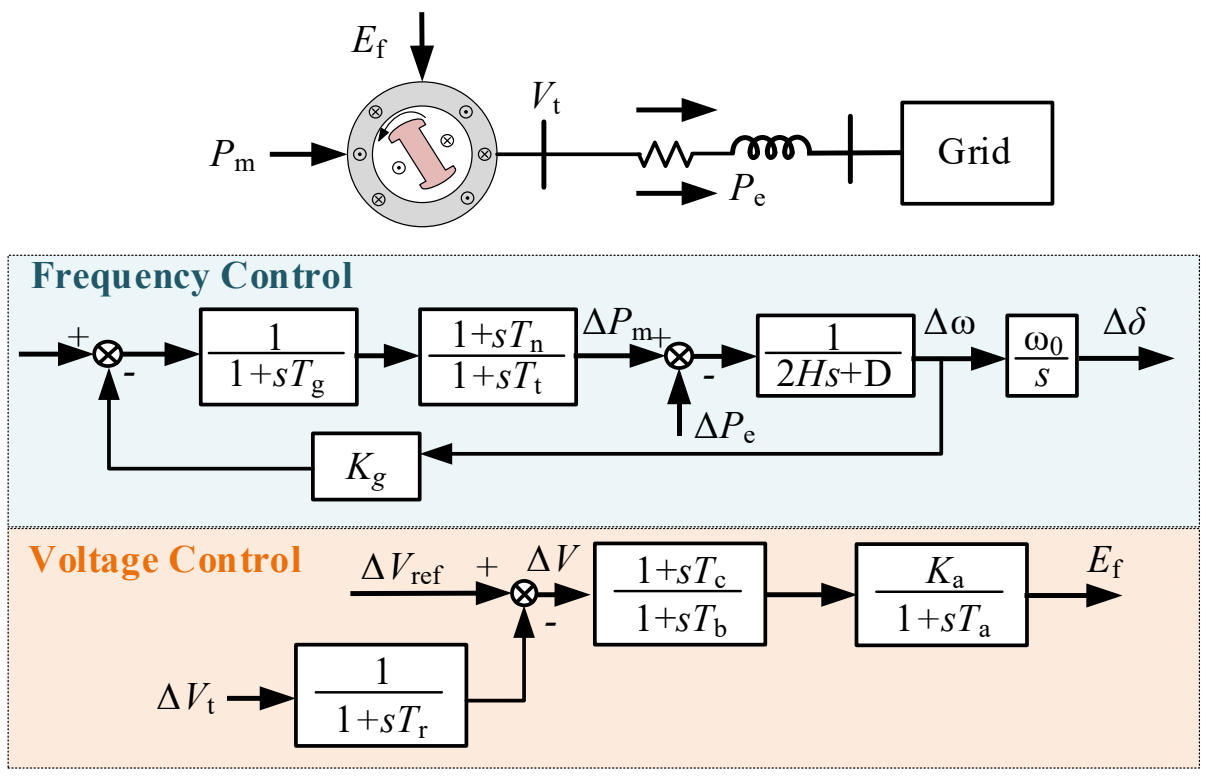

**Fig. 16.** SG dynamic model and control structure.

TABLE III
CONTROL PARAMETERS OF SG

| Parameter | Symbol | Value |
|---|---|---|
| Governor gain | $K_g$ | 20 |
| Governor lead time constant | $T_n$ | 2.1 s |
| Governor lag time constant | $T_t$ | 7 s |
| Governor lag time constant | $T_g$ | 0.2 s |
| AVR gain | $K_a$ | 100 |
| Voltage-measurement time constant | $T_r$ | 0.01 s |
| AVR time constant | $T_a$ | 0.015 s |
| AVR lag time constant | $T_b$ | 10 s |
| AVR lead time constant | $T_c$ | 1 s |

The governor, excitation, and rotor-angle transfer functions are

$$\begin{cases} G_g(s) = \frac{K_g(1+T_n s)}{(1+T_t s)(1+T_g s)} \\ G_A(s) = \frac{K_a(1+T_c s)}{(1+T_r s)(1+T_a s)(1+T_b s)} \\ G_r(s) = \frac{\omega_0}{s[2Hs+D+G_g(s)]} \end{cases} \tag{26}$$

where $H$ is the inertia constant, $D$ is the damping coefficient, and $\omega_0$ is the rated electrical angular frequency. The remaining control parameters are listed in Table III.

The $d$- and $q$-axis reactance functions and the field transfer function are

$$\begin{cases} X_d(s) = X_d'' + \frac{X_d'-X_d''}{1+T_{d0}''s} + \frac{X_d-X_d'}{(1+T_{d0}'s)(1+T_{d0}''s)} \\ X_q(s) = X_q'' + \frac{X_q'-X_q''}{1+T_{q0}''s} + \frac{X_q-X_q'}{(1+T_{q0}'s)(1+T_{q0}''s)} \\ F_d(s) = [(1+T_{d0}'s)(1+T_{d0}''s)]^{-1} \end{cases} \tag{27}$$

where $X_d$ and $X_q$ are the synchronous reactances. Primes and double primes denote transient and subtransient quantities, respectively. The associated $T$ parameters are open-circuit time constants. The machine parameters are based on the IEEE 39-bus generator data in [31].

Let $\alpha$ denote the rotor-to-terminal angle difference. The linearized stator equations can be written as $\boldsymbol{A}_{\mathrm{SG}}[\Delta\alpha, \Delta v]^T = \boldsymbol{B}_{\mathrm{SG}}[\Delta P, \Delta Q]^T$, where the coefficient matrices are given by

$$\begin{cases} \boldsymbol{A}_{\mathrm{SG}} = \begin{bmatrix} V_{q0} + X_q I_{d0} & V_{d0} + X_q I_{q0} \\ -V_{d0} + X_d I_{q0} & V_{q0} - X_d I_{d0} + F_d G_A V_{t0} \end{bmatrix} \\ \boldsymbol{B}_{\mathrm{SG}} = \frac{1}{V_{t0}} \begin{bmatrix} X_q \cos\alpha_0 & -X_q \sin\alpha_0 \\ -X_d \sin\alpha_0 & -X_d \cos\alpha_0 \end{bmatrix} \end{cases} \tag{28}$$

where $V_{d0}, V_{q0}, I_{d0}, I_{q0}$ are the operating-point stator components, and $V_{t0}$ is the terminal-voltage magnitude.

**2) Grid-forming converter:** The GFM implementation uses the virtual synchronous control structure in Fig.17, with parameters listed in Table IV. The active-power loop governs frequency dynamics, while the reactive-power loop regulates the internal voltage magnitude.

The filter in (5) is $F_p(s) = (1+\tau_p s)^{-1}$. The parameters in Fig. 17 satisfy $J_{\mathrm{vsg}} = 2HS_r/\omega_0^2$ and $D_{\mathrm{vsg}}^{\mathrm{pu}} = \omega_0 D_{\mathrm{vsg}}/S_r$. The port voltage $V_{t0}$ is evaluated at the controlled-voltage terminal; external transformer and line coupling are represented in the network model.

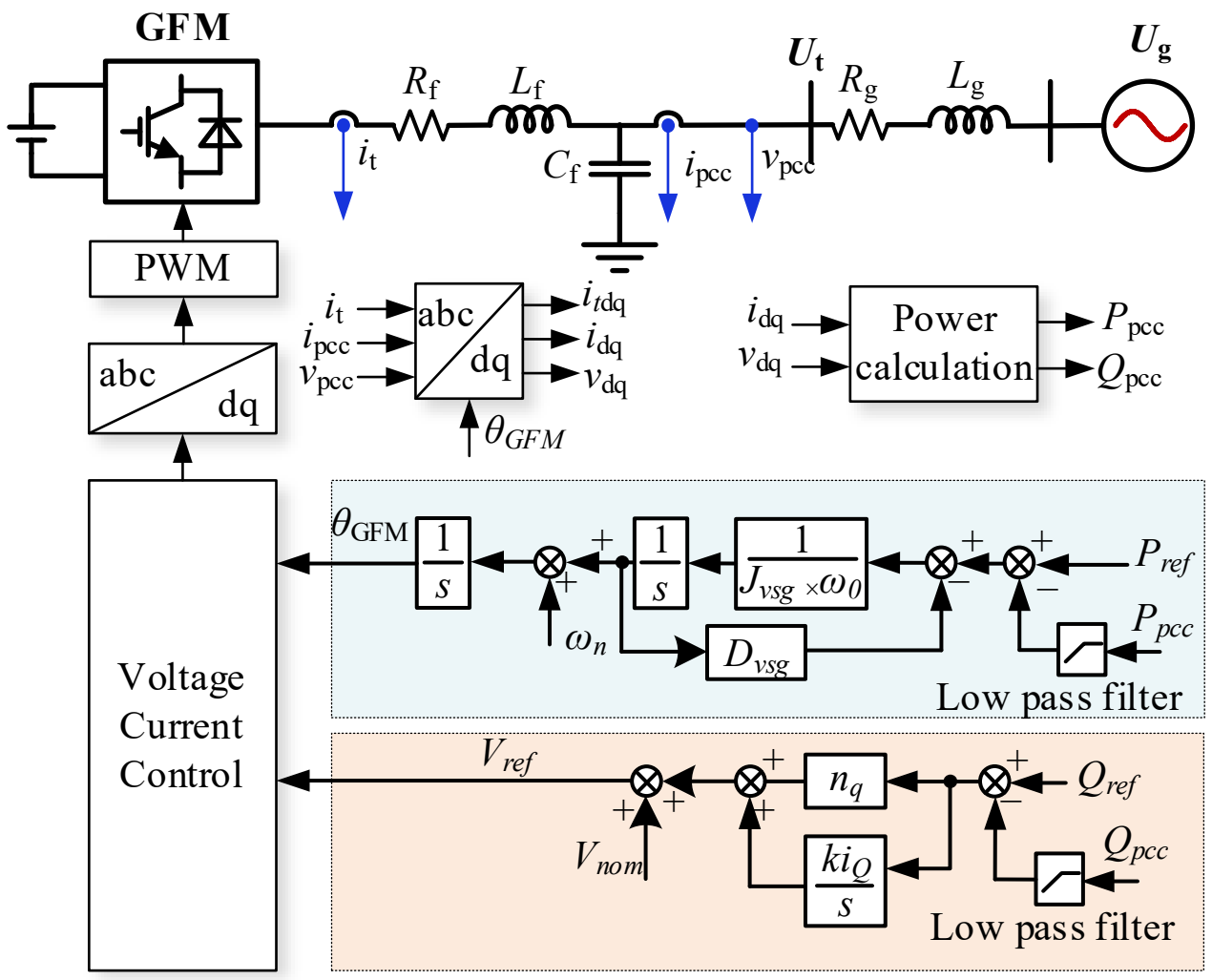

**Fig. 17.** GFM model and control structure.

TABLE IV
GFM CONTROL PARAMETERS

| Parameter | Symbol | Value |
|---|---|---|
| Virtual inertia constant | $H$ | 3 s |
| Normalized damping coefficient | $D_{\mathrm{vsg}}^{\mathrm{pu}}$ | 20 |
| Power-measurement time constant | $\tau_p$ | 0.001 s |
| Reactive-power proportional gain | $n_q$ | $1.5\times10^{-5}$V/var |
| Reactive-power integral gain | $K_{i,Q}$ | $1.0\times10^{-4}$V/(var·s) |
| Voltage-loop PI gains | $K_{p,v}$,$K_{i,v}$ | 0.01, 0.5 |
| Current-loop PI gains | $K_{p,i}$,$K_{i,i}$ | 0.5, 5 |

**3) Grid-following DFIG:** The DFIG structure is shown in Fig. 18. Control parameters are listed in Table V, while rated and electrical parameters follow Table I of [18].

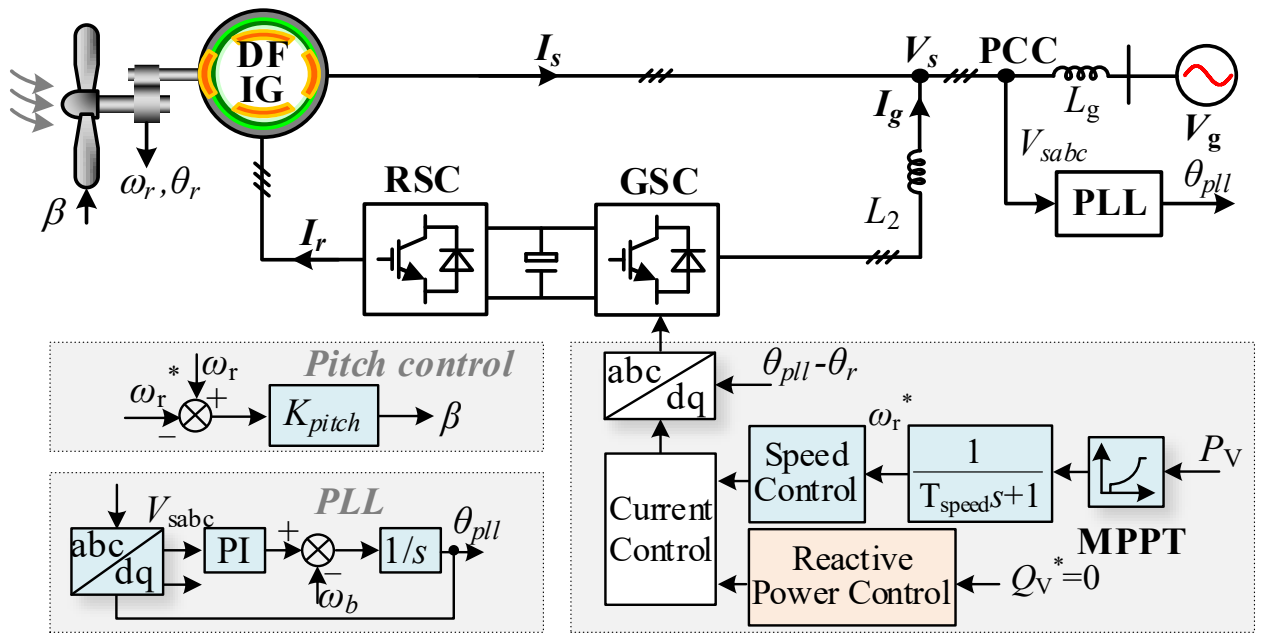

**Fig. 18.** GFL-DFIG model and control structure.

TABLE V
DFIG CONTROL PARAMETERS

| Parameter | Symbol | Value |
|---|---|---|
| PLL PI gains | $K_{p,\mathrm{PLL}}$, $K_{i,\mathrm{PLL}}$ | 60, 1400 |
| Speed-loop PI gains | $K_{p,\omega}$, $K_{i,\omega}$ | 1.5, 0.5 |
| Reactive-power-loop PI gains | $K_{p,Q}$, $K_{i,Q}$ | 0.3, 20 |
| Rotor-current-loop PI gains | $K_{p,c}$, $K_{i,c}$ | 10, 2 |
| Speed-reference time constant | $T_{\mathrm{speed}}$ | 5 s |
| Pitch-control gain | $K_{\mathrm{pitch}}$ | 150 |
| Pitch-actuator time constant | $T_{\mathrm{pitch}}$ | 0.01 s |

Define the speed-loop, reactive-power-loop, and PLL transfer functions as

$$\begin{cases} C_\omega(s) = K_{p,\omega} + \frac{K_{i,\omega}}{s}, C_Q(s) = K_{p,Q} + \frac{K_{i,Q}}{s} \\ G_{\mathrm{PLL}}(s) = \frac{V_0(K_{p,\mathrm{PLL}}s + K_{i,\mathrm{PLL}})}{s^2 + V_0 K_{p,\mathrm{PLL}}s + V_0 K_{i,\mathrm{PLL}}} \end{cases} \tag{29}$$

where $V_0$ is the operating-point voltage and the controller gains are listed in Table V.

The speed-reference transfer function is $R_\omega(s) = K_r/(1 + T_{\mathrm{speed}}s)$, where $K_r$ is the speed-reference gain. The transfer function $G_p(s)$ is defined by $\Delta\omega_r - \Delta\omega_{r,\mathrm{ref}} = -G_p(s)\Delta P$. It is derived from the speed and pitch-control dynamics shown in Fig. 18, following the model in [18].

The denominator functions in (6) are

$$\begin{cases} \mathcal{D}_P(s) = 1 + \frac{\omega_{r0} X_m V_0}{X_s} C_\omega G_p - \frac{P_0}{\omega_{r0}}\left(R_\omega - G_p\right) \\ \mathcal{D}_Q(s) = 1 + \frac{X_m V_0}{X_s} C_Q \end{cases} \tag{30}$$

The reactances $X_m, X_s$ are the magnetizing and total stator reactances, respectively. The quantities $\omega_{r0}$ and $P_0$ denote the operating-point rotor speed and active power.